\documentclass[aps, prl,twocolumn,longbibliography,superscriptaddress,floatfix]{revtex4}
\usepackage[latin9]{inputenc}
\usepackage{amssymb}
\usepackage{graphicx}
\usepackage{amsmath}
\usepackage{color}
\usepackage{mathrsfs}
\usepackage{float}
\usepackage{indentfirst}
\usepackage{mathrsfs}
\usepackage{float}
\usepackage{indentfirst}
\usepackage{textcomp}
\usepackage{comment}
\usepackage{mathtools}
\usepackage{natbib,hyperref}
\usepackage{soul}

\begin{document}

\title{Numerical study of bi-layer two-orbital model for La$_{3}$Ni$_{2}$O$_{7}$ on a plaquette ladder}
\author{Yang Shen}
\affiliation{Key Laboratory of Artificial Structures and Quantum Control, School of
Physics and Astronomy, Shanghai Jiao Tong University, Shanghai 200240, China}
\author{Jiale Huang}
\affiliation{Key Laboratory of Artificial Structures and Quantum Control, School of
Physics and Astronomy, Shanghai Jiao Tong University, Shanghai 200240, China}
\author{Xiangjian Qian}
\affiliation{Key Laboratory of Artificial Structures and Quantum Control, School of
Physics and Astronomy, Shanghai Jiao Tong University, Shanghai 200240, China}
\author{Guang-Ming Zhang}
\email{gmzhang@tsinghua.edu.cn}
\affiliation{State Key Laboratory of Low-Dimensional Quantum Physics and Department of
Physics, Tsinghua University, Beijing 100084, China}
\affiliation{Frontier Science Center for Quantum Information, Beijing 100084, China}
\affiliation{School of Physical Science and Technology, ShanghaiTech University, Shanghai 201210, China}
\author{Mingpu Qin}
\email{qinmingpu@sjtu.edu.cn}
\affiliation{Key Laboratory of Artificial Structures and Quantum Control, School of
Physics and Astronomy, Shanghai Jiao Tong University, Shanghai 200240, China}
\affiliation{Hefei National Laboratory, Hefei 230088, China}
\date{\today}
 
\begin{abstract}
The recently discovered high-$T_c$ superconductivity in La$_{3}$Ni$_{2}$O$_{7}$ with $T_c \approx 80K$ provides another intriguing platform to explore the microscopic mechanism of unconventional superconductivity. In this work, we study a previously proposed bi-layer two-orbital model Hamiltonian for La$_{3}$Ni$_{2}$O$_{7}$ [Y. Shen, et al, Chinese Physics Letters 40, 127401 (2023)] on a plaquette ladder, which is a minimum setup with two-dimensional characteristic. We employ large-scale Density Matrix Renormalization Group calculations to accurately determine the ground state of the model. We determine the density, magnetic structure, and the pairing property of the model. We find that with large effective inter-layer anti-ferromagnetic exchange for the 3$d_{z^2}$ orbital, both spin, charge, and pairing correlation display quasi-long-range behavior, which could be viewed as a precursor of possible true long-range order in the two dimensional limit. Interestingly, sign oscillation for the pairing correlation are observed for both the 3$d_{x^2-y^2}$ and 3$d_{z^2}$ orbitals, indicating the presence of possible pair density wave in the system. Even though we only study the model on a quasi one-dimensional plaquette ladder geometry due to the computational difficulty, the results on the spin, charge, and pairing correlation provide valuable insight in the clarification of the properties of La$_{3}$Ni$_{2}$O$_{7}$ in the future.
\end{abstract}

\maketitle
\textbf{Introduction}
Unveiling the microscopic mechanism of unconventional superconductivity is one of the main challenges in modern condensed matter physics. Recently, another series of unconventional superconductors other than cuprates \cite{bednorz1986possible} and Iron-based superconductors \cite{kamihara2008iron,kamihara2006iron} are synthesized. The infinite-layer nickelates, initially proposed as $3d^9$ analogs to cuprates in 1999 \cite{PhysRevB.59.7901}, is found to be superconducting in epitaxial thin films twenty years later \cite{li2019superconductivity,osada2020superconducting,zeng2022superconductivity,osada2021nickelate}. While infinite-layer nickelates have active 3$d_{x^2-y^2}$ orbital degrees of freedom on a two-dimensional square lattice, similar as in cuprates, electronic band differences in self-doping effect and $p-d$ energy level splitting are evident \cite{PhysRevB.70.165109,nomura2022superconductivity}, making it fit more into the Mott-Hubbard side than the charger transfer side \cite{doi:10.1073/pnas.2007683118}. However, infinite-layer nickelates do not behave like antiferromagnetic Mott insulators. The resistivity exhibits metallic behavior at high temperatures and a low-temperature upturn \cite{li2019superconductivity,PhysRevB.101.020501} without long-range magnetic order \cite{HAYWARD2003839} in the parent compounds.

\begin{figure}[tbp]
\includegraphics[width=0.5\textwidth]{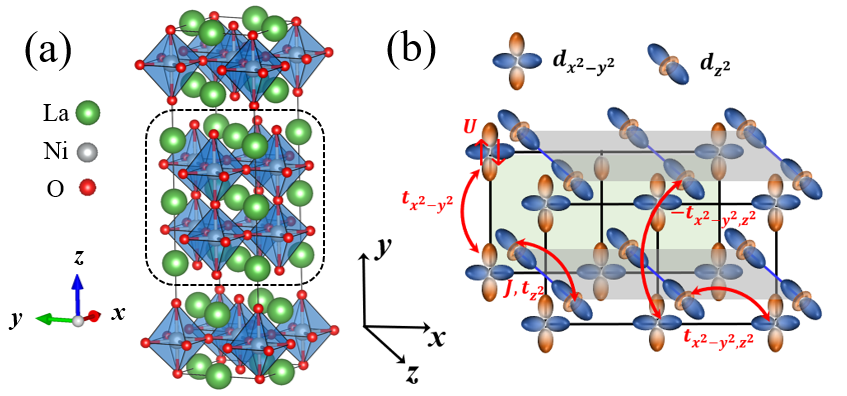}
\caption{(a) Crystal structure and (b) schematic illustration of the bi-layer double-orbital (containing $3d_{x^2-y^2}$ and $3d_{z^2}$) model for high-$T_c$ nickelate superconductor La$_{3}$Ni$_{2}$O$_{7}$. The dashed frame in (a) encricles the bilayer oxygen octahedral structure. In our model, the inter-layer hopping of 3$d_{z^2}$ orbital $t_{z^2}$ is set as the energy unit, the intra-layer hopping between 3$d_{x^2-y^2}$ orbitals $t_{x^{2}-y^{2}}=0.8$ and the intra-layer hybridization between the 3$d_{z^2}$ and 3$d_{x^2-y^2}$ orbitals $t_{x^{2}-y^{2},z^{2}}=0.4$. Notice that the sign of intra-layer hopping $t_{x^{2}-y^{2},z^{2}}$ along the vertical and horizontal direction are opposite due to the spatial symmetries of two $e_g$ orbitals. The super- exchange interaction between the inter-layer $3d_{z^2}$ orbitals is chosen as $J = 0.5$. The Hubbard repulsion $U$ within the $3d_{x^{2}-y^{2}}$ orbitals is 8. We neglect the intra-layer hopping of the $3d_{z^2}$ orbital electrons, the inter-layer hopping of the $3d_{x^2-y^2}$ orbital electrons and the inter-layer hybridization between the 3$d_{z^2}$ and 3$d_{x^2-y^2}$ orbitals because they are small \cite{PhysRevLett.131.126001}.}
\label{model}
\end{figure}

The recent discovery of superconductivity in pressurized La$_3$Ni$_2$O$_7$ with $T_c$ around 80 K has attracted a lot of intention for the nickelate family \cite{sun2023signatures,Hou_2023,zhang2024high,ZHANG2024147,PhysRevX.14.011040,Wang_2024}. Shortly after the discovery of La$_3$Ni$_2$O$_7$, La$_4$Ni$_3$O$_{10}$ exhibiting superconductivity with $T_c \approx 30$ K under pressure was also reported \cite{zhu2024superconductivity,PhysRevB.109.165140,PhysRevB.109.144511,li2024signature,li2024structural,zhang2023superconductivity,PhysRevLett.133.136001}. Similar to the infinite-layer nickelates, these newly-discovered Ruddlesden-Popper phase nickelates are layered materials composed of alternating charge reservoir planes and NiO planes. The primary structural distinction is the number of linked NiO layers through apical oxygens, which subsequently influences the nickel's valence electron count. For La$_3$Ni$_2$O$_7$, every two adjacent NiO bilayer perovskite structure (designated as 2222, another structural candidate is 1313 \cite{abadi2024electronic,yang2024orbital}) is interconnected by apical oxygens along the stacking direction (Fig.~\ref{model}(a)).
 Density Functional Theory calculations indicate that both the 3$d_{x^2-y^2}$ and 3$d_{z^2}$ orbitals of Ni cations are strongly hybridized with the oxygen 2$p$ orbitals near the Fermi surface at high pressure \cite{sun2023signatures,zhang2024structural,PhysRevB.108.L180510,PhysRevB.108.L201121,PhysRevB.83.245128,gu2023effective}. The intra-layer 3$d_{x^2-y^2}$ electrons on the NiO$_2$ plane exhibit similarities to cuprates. However, in La$_3$Ni$_2$O$_7$ the 3$d_{z^2}$ orbitals are also relevant in the high pressure phases, which are nearly flat and experience a significant inter-layer coupling through the apical oxygen, due to the quantum confinement of the NiO$_2$ bilayer structure. Consequently, the energy splitting of the Ni cations leads to a substantial alteration in the distribution of the averaged valence state, resulting in a nominal Ni 3$d^{7.5}$ electronic configuration. Therefore, a minimum two orbitals within $e_g$ multiples are necessary for modeling La$_3$Ni$_2$O$_7$ \cite{kaneko2024pair,shen2023effective,PhysRevLett.131.126001,zhang2024structural,PhysRevLett.132.146002}. Various theoretical studies have carefully revealed the roles of Hund coupling \cite{PhysRevB.108.L180510,PhysRevB.109.L081105,PhysRevLett.132.146002,PhysRevB.109.115114,PhysRevB.108.174511}, inter-layer exchange interaction \cite{shen2023effective,PhysRevB.110.024514,PhysRevLett.132.036502,PhysRevB.108.L140505,gu2023effective,Xue_2024}, electronic correlation \cite{PhysRevB.108.L201121,wu2023charge,liu2024electronic} and electron-phonon coupling \cite{zhan2024cooperation,ouyang2024absence} in realizing the high temperature superconductivity, and  the charge and spin instability \cite{PhysRevLett.131.206501,PhysRevB.108.L180510,PhysRevB.108.125105,chen2023critical,PhysRevLett.132.256503,dan2024spin,zhou2024electronic,khasanov2402pressure} were also recently discussed. 
 
The Fermi surface for the high pressure phase from band structure calculation reveals the presence of one electron-like $\alpha$ pocket and another large hole-like $\beta$ pocket from strong $e_g$ multiple hybridization, which was also confirmed by the ARPES measurement \cite{abadi2024electronic,yang2024orbital}. Band structure calculation also show  the appearance of a small hole-like $\gamma$ pocket mainly contributed by the $3d_{z^2}$ in the high pressure phase \cite{sun2023signatures,zhang2024structural,PhysRevB.108.L180510,PhysRevB.108.L201121,PhysRevB.83.245128,gu2023effective}. The pairing symmetry in La$_3$Ni$_2$O$_7$ is widely studied by effective Hamiltonian approaches based on this Fermi surface topology. Theoretical models that emphasize the $\gamma$ band typically predict an $s_{\pm}$-wave pairing with sign-reversal gaps due to the coupling between the $\gamma$ band and other bands \cite{PhysRevB.108.L140505,gu2023effective,luo2024high,PhysRevLett.131.236002}. In these models, Cooper pairs are pre-formed on the interlayer 3$d_{z^2}$ valence bonds and then gain phase coherence within the NiO$_2$ planes through the inter-orbital hybridization, which leads to superconductivity \cite{shen2023effective,PhysRevB.108.L201108}. On the other hand, the pairing interaction in the $\alpha$ and $\beta$ bands suggests a $d$-wave character of pairing symmetry \cite{luo2024high,liu2023role,Jiang_2024,PhysRevB.110.024514}, as Ni $3d_{x^2-y^2}$ and O $2p$ electrons (mainly from the $\beta$ bands) dominate the low-energy density of states around the Fermi level. This assumption relies on that the doping level of superconducting phase resembles the optimal doping in the cuprates, if one take $\gamma$ band's role as hole dopants into other bands. Furthermore, the $s$-wave pairing trend of the $\alpha$ band \cite{PhysRevB.110.024514} and the additional interlayer pairing interaction between Ni $3d_{x^2-y^2}$ electrons, induced by Hund's coupling \cite{PhysRevLett.132.146002,PhysRevB.109.115114}, have also been demonstrated. 

In this work, we revisit the bi-layer two-orbital model proposed by us in \cite{shen2023effective}. In \cite{shen2023effective}, we study a minimal one-dimensional setup and found strong pairing tendency in both orbitals. Here we study the model on a more realistic plaquette ladder geometry (see Fig.\ref{model} (b)), which is a minimum setup with two-dimensional characteristic. We investigate the model using large scale Density Matrix Renormalization Group (DMRG) calculations. We discover that with a substantial effective inter-layer antiferromagnetic exchange for the $3d_{z^2}$ orbital, there is a manifestation of quasi-long-range behavior in spin, charge, and pairing correlations.  The charge order exhibits pronounced orbital selectivity, with the $3d_{x^{2}-y^{2}}$ orbital exclusively manifesting a quasi-long-range charge density wave order with a period of 2 lattice units. The Neel-type antiferromagnetic instabilities are present for both the $3d_{x^{2}-y^{2}}$ and $3d_{z^{2}}$ orbitals. Moreover, sign oscillation for the pairing correlation are observed for
both the $3d_{x^2-y^2}$ and $3d_{z^2}$ orbitals, indicating the presence of possible pair density wave in the system. Although our study is confined to a plaquette ladder rather than a genuine two-dimensional lattice due to computational constraints, the results on pairing correlation provide valuable insight in the clarification of charge, spin and pairing properties in La$_{3}$Ni$_{2}$O$_{7}$ in the future.

\textbf{Effective model Hamiltonian and method}. - The bi-layer two-orbital Hamiltonian of square planars coordinated Ni cations consisting of the $3d_{z^{2}}$ and $3d_{x^{2}-y^{2}}$ orbitals \cite{shen2023effective} is,
\begin{eqnarray}
H &=&-t_{x^{2}-y^{2}}\sum_{\langle ij\rangle ,\sigma ,a=1,2}(c_{a,i,\sigma
}^{\dagger }c_{a,j,\sigma }+h.c)  \notag \\
&&-\mu _{x^{2}-y^{2}}\sum_{i,a=1,2}n_{a,i}^{c}+U\sum_{i,a=1,2}n_{a,i\uparrow
}^{c}n_{a,i\downarrow }^{c}  \notag \\
&&-t_{x^{2}-y^{2},z^{2}}\sum_{i,\sigma ,a=1,2}(d_{a,i,\sigma }^{\dagger }%
\widetilde{c}_{a,i,\sigma }+h.c)  \notag \\
&&-t_{z^{2}}\sum_{i,\sigma }(d_{1,i,\sigma }^{\dagger }d_{2,i,\sigma }+h.c)
\notag \\
&+&J\sum_{i}\mathbf{S}_{1,i}^{{d}}\cdot \mathbf{S}_{2,i}^{{d}}-\mu
_{z^{2}}\sum_{i,a}n_{a,i}^{d}  
\label{ham}
\end{eqnarray}%

where \( c_{a,i,\sigma}^{\dagger} \) (\( d_{a,i,\sigma}^{\dagger} \)) creates a \( 3d_{x^{2}-y^{2}} \) (\( 3d_{z^{2}} \)) electron at site \( i \) on layer \( a = 1,2 \). 
The operator \( \widetilde{c}_{a,i,\sigma} \) is defined as
\begin{equation}
\widetilde{c}_{a,i,\sigma} = \frac{1}{2} \left( c_{a,i+x,\sigma} + c_{a,i-x,\sigma} - c_{a,i+y,\sigma} - c_{a,i-y,\sigma} \right)
\end{equation}
\( t_{z^{2}} \) represents the hopping integral for the \( 3d_{z^{2}} \) electrons between the two layers and is taken as the energy unit. 
\( t_{x^{2}-y^{2}} \) is the hopping integral for the \( 3d_{x^{2}-y^{2}} \) electrons within each layer.  The term \( t_{x^{2}-y^{2},z^{2}} \) denotes the intra-layer hybridization between the \( 3d_{x^{2}-y^{2}} \) electrons and \( 3d_{z^{2}} \) electrons of nearest neighbors, with opposite signs for vertical and horizontal directions \cite{PhysRevLett.131.126001}, as can be seen from the definition of $\widetilde{c}_{a,i,\sigma}$. The
Hubbard repulsion $U$ is invoked within the $3d_{x^{2}-y^{2}}$ orbitals. The chemical potentials \( \mu_{x^{2}-y^{2}} \) and \( \mu_{z^{2}} \) correspond to the \( 3d_{x^{2}-y^{2}} \) and \( 3d_{z^{2}} \) orbitals of nickel, respectively. We choose \( \mu_{z^{2}} \) that is significantly smaller than \( \mu_{x^{2}-y^{2}} \) to ensure the \( 3d_{z^{2}} \) orbital is near half-filling, while the \( 3d_{x^{2}-y^{2}} \) orbital is close to quarter-filling. It was shown that there is a large inter-layer anti-ferromagnetic super-exchange for the $3d_{z^{2}}$ orbitals via apical oxygen anions \cite{shen2023effective} which is the main cause of pairing instability. {Thus we also consider the anti-ferromagnetic exchange interaction $J$ between these
these orbitals by prohibiting double occupancy of the \( 3d_{z^{2}} \) orbital, $i.e.$, imposing the local constraint \( n_{a,i}^{d} = n_{a,i,\uparrow}^{d} + n_{a,i,\downarrow}^{d} < 2\).}

The crucial role of the interlayer exchange coupling are highlighted by the emergence of the $d_{z^2}$ orbitals crossing the Fermi level and the flattening of the apical Ni-O-Ni bonding angle during the structural transition into the superconducting phase under pressure. In this model, we have neglected the intra-layer (inter-layer) hopping for the \( 3d_{z^{2}} \) (\( 3d_{x^{2}-y^{2}} \)) orbitals since they are tiny compared with the above parameters. 

In order to explore the low-energy physics of the effective model given by Eq.~(\ref{ham}),
we employ the Density Matrix Renormalization Group method \cite{PhysRevLett.69.2863,PhysRevB.48.10345} to numerically solve the model on a plaquette ladder that captures the double-layer structure with a length $L_x=32$ as depicted in Fig.~\ref{model}(b).
We select the hopping parameter $t_{z^{2}}$ as our energy unit, setting $t_{x^{2}-y^{2}} = 0.8$ and $t_{x^{2}-y^{2},z^{2}} = 0.4$ \cite{PhysRevB.83.245128}. The
Hubbard repulsion $U$ within the $3d_{x^{2}-y^{2}}$ orbitals is 8. Our focus is on the large $J$ limit, hence the local antiferromagnetic spin coupling is set to $J = 0.5$.
Under an averaged filling of $n = 3/4$, we perform a scan of $\Delta_{\mu} = \mu_{x^{2}-y^{2}} - \mu_{z^{2}}$ to ensure that the $3d_{z^{2}}$ orbitals are at $1/16$ hole doping away from half-filling, while the $3d_{x^{2}-y^{2}}$ orbital electrons exhibit a filling of $9/16$. With these parameters, the $3d_{z^{2}}$ orbital electrons are close to the Mott-insulating limit, whereas the $3d_{x^{2}-y^{2}}$ orbital electrons are in a regime of large doping.
In our DMRG calculations, we push the bond dimension up to $m = 40000$ with a truncation error $\epsilon \approx 1 \times 10^{-5}$. Extrapolations with truncation error are also performed to reduce the error caused by finite bond dimension.

\begin{figure}[t]
\includegraphics[width=0.5\textwidth]{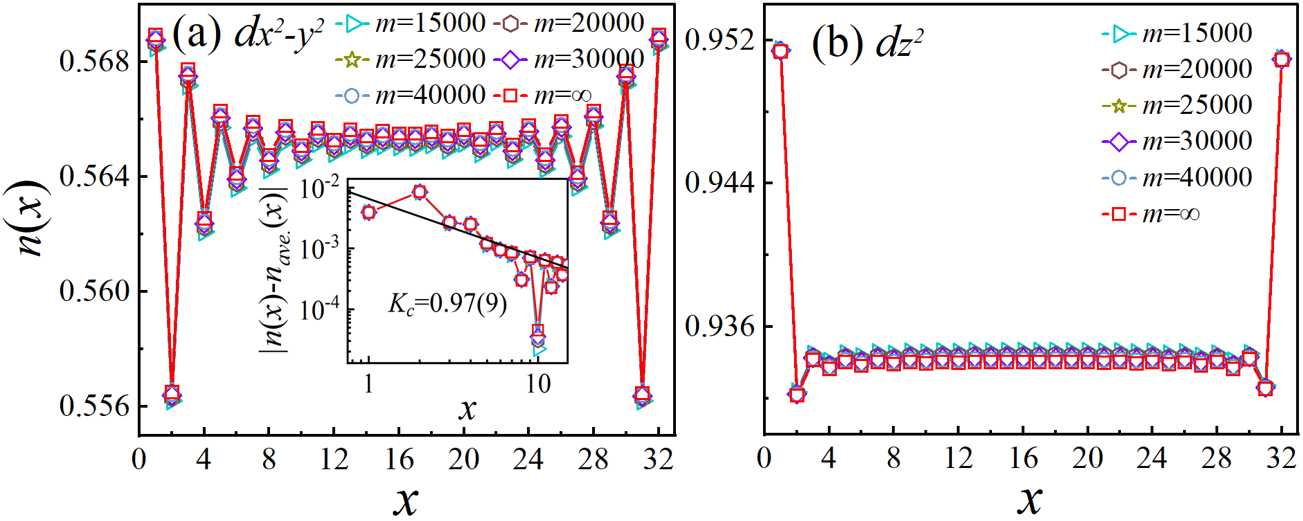}
\caption{ The charge density distribution of the (a) $3d_{x^2-y^2}$ and (b) $%
3d_{z^2}$ orbitals. The chemical potential difference $\Delta_{\protect\mu}$
is $1.2$ to target the $1/16$ hole doping level on the $3d_{z^2}$
orbitals and $9/16$ electron filling on the $3d_{x^2-y^2}$ orbital. Both results of finite kept states and extrapolation are plotted to show the good convergence of charge density in DMRG calculations. The inset in (a) depicts the power-law decay of charge density on the $3d_{x^2-y^2}$ orbital (utilizing solely the early segment of extrapolated envelope data after the average density has been subtracted), indicating a quasi-long-range charge order. The charge density wave on the 3$d_{x^2-y^2}$ orbitals has a wavelength of $2$ lattice units. The relatively uniform charge density patterns away from the boundary can be seen on the $3d_{z^2}$ orbitals.}
\label{charge}
\end{figure}

\textbf{Charge and spin density distributions}. - Fig.~\ref{charge} shows the local charge density profiles of both the 3$d_{z^2}$ and 3$d_{x^2-y^2}$ electrons, where rung charge density is defined as $n(x)=\sum_{y=1}^{L_{y}}\left\langle\hat{n}_{i}(x,y)\right\rangle / L_{y}$ in the width direction along which charge density distributions are identical. We display both finite kept states results and the extrapolation with kept states results to denote the convergence of charge density in the DMRG calculations. The charge density of the 3$d_{x^2-y^2}$ electrons (Fig.~\ref{charge} (a)) exhibits charge density wave (CDW) structure with wavelength of $2$ lattice units, characterizing a $(\pm \pi, 0)$ vertical charge stripe. In the inset, we show a simple polynomial fit ($\sim x^{-K_c}$) of the amplitude of the CDW, which gives $K_c=0.97(9)$ and suggests a strong CDW tendency. The modulation of the charge density of 3$d_{z^2}$ orbital is less pronounced, maintaining a constant amplitude in the bulk. Thus a CDW instability is expected to occur in the $x-y$ plane for the 3$d_{x^2-y^2}$ orbital, while the 3$d_{z^2}$ orbitals shows no evidence of ordering tendency. Since 3$d_{x^2-y^2}$ orbital is nearly one-quarter filling, the CDW on 3$d_{x^2-y^2}$ orbitals is similar to nearly $1/2$ hole doped Hubbard ladder, which may be categorized into Tomonaga-Luttinger liquid phase with dominant 2$k_F$ density oscillation mode \cite{giamarchi2003quantum,PhysRevB.107.125114}. Though the findings from the plaquette ladder foresee distinct CDW attributes as opposed to the one-dimensional setup in which case both 3$d_{x^2-y^2}$ and 3$d_{z^2}$ orbitals exhibit (quasi) long-range CDW with varying wavelengths \cite{shen2023effective}, these models collectively encapsulate the quasi-one-dimensional nature deemed essential for the emergence of CDW \cite{seo1996electronic}. It is also interesting to notice that an incommensurate  
diagonal CDW at $(\pm \pi, \delta)$ $k$-vector (with a small $\delta$) was previously proposed in the high pressure phase, with charge modulation on the 3$d_{z^2}$ channels\cite{PhysRevLett.131.206501}. 
The involvement of different active orbitals indicates a fundamentally different origin for the potential CDW instability and need an experimental confirmation. Moreover, the CDW pattern, though similarly carries 3$d_{x^2-y^2}$ activeness, also distinguishes La$_3$Ni$_2$O$_7$ from the cuprates, where the charge order is in the form of insulating stripe order intertwined with staggered magnetic correlations. 

\begin{figure}[t]
\includegraphics[width=0.5\textwidth]{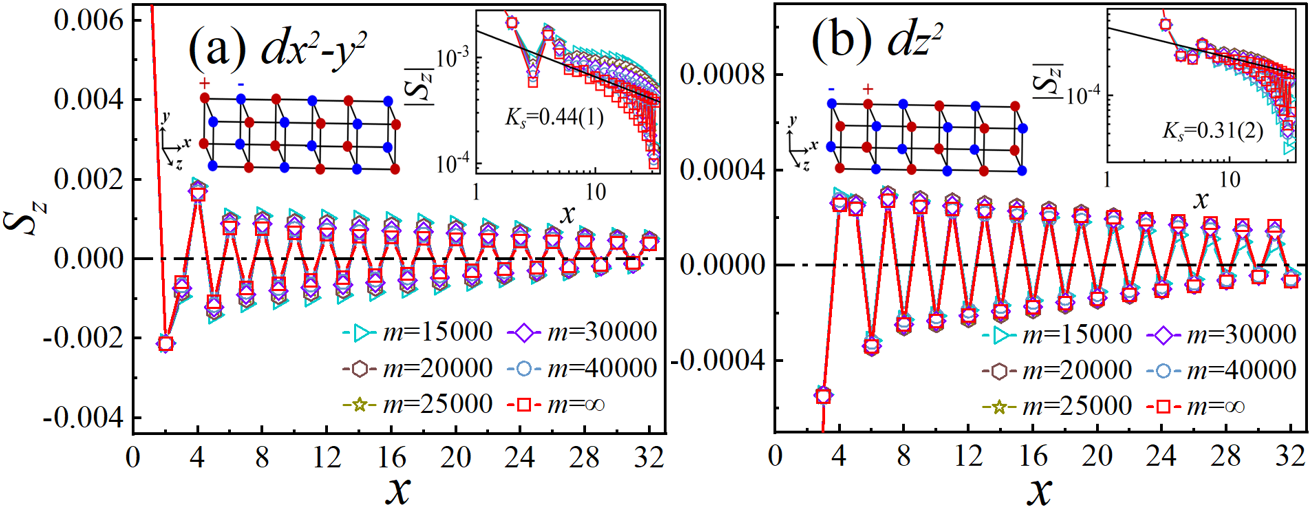}
\caption{ The spin density distribution of the  (a) $3d_{x^2-y^2}$ and (b) $3d_{z^2}$ orbitals. Both results of finite kept states and extrapolation are plotted to show the good convergence of the DMRG calculations. The spin density wave patterns with tiny amplitude and oscillated wavelength 2 can be seen on both $3d_{z^2}$ and $3d_{x^2-y^2}$ orbitals. The insets display the power-law decay behavior of absolute spin density amplitudes. The dash-dotted horizontal lines represent zero amplitude of spin density. The sign structures of spin density are also displayed in (a) and (b). Within each layer, spin density on $3d_{x^2-y^2}$ orbitals displays Neel-type AFM correlations along the x-axis and ferromagnetic along the y-axis. The $3d_{z^2}$ spin densities on adjacent inter-layer atomic sites exhibit opposite signs, and these  orbitals on the same atomic site have opposite spin signs.}
\label{spin}
\end{figure}

The local spin densities of both the $3d_{x^2-y^2}$ and $3d_{z^2}$ orbitals are displayed in Fig.~\ref{spin} (a) and (b), as defined by $\langle \hat{S}_i^z \rangle$. To avoid directly calculating the more demanding spin-spin correlations in the DMRG method, we apply a symmetry breaking pinning magnetic field with the strength $%
h_m = 0.5$ at one site in the left edge, and probe the magnetic structure by calculating the local spin density \cite{PhysRevLett.99.127004}. Both orbitals show Neel type spin density wave (SDW) pattern, but magnitudes on the $3d_{x^2-y^2}$ are significantly larger than those on the $3d_{z^2}$ orbitals. Polynomial fits ($\sim x^{-K_s}$) give exponents $K_s = 0.44(1)$ and $0.31(2)$ for the $3d_{x^2-y^2}$ and the $3d_{z^2}$ orbitals respectively, indicating the quasi-long-range order and the absence of spin excitation gap. The stronger SDW on $3d_{x^2-y^2}$ orbitals resembles the characteristics of the nearly $1/2$ hole doped Hubbard ladder in Tomonaga-Luttinger liquid phase. For both orbitals, the spin densities in each plaquette with same $x$ coordinate have the same absolute magnitude, and the corresponding sign structures are displayed in the insets of Fig.~\ref{spin} (a) and (b). The inter-layer atmoic sites have opposite sign, which can be attributed to the larger inter-layer AFM coupling $J$ for the $3d_{z^2}$ orbitals. Within each layer, the spin density exhibits Neel-type antiferromagnetic correlations along the $x$-direction, contrasting with ferromagnetic correlations along the $y$-direction. Experimental observations have not yet directly pinpointed an SDW instability for the high-pressure phase. An SDW order near the wave vector \( Q = (\pi/2, \pi/2) \) at ambient pressure has been inferred from resonant inelastic X-ray scattering \cite{chen2024electronic}, muon spin rotation\cite{PhysRevLett.132.256503,khasanov2024pressure}, and nuclear magnetic resonance \cite{dan2024spin} measurements. However, our in-plane magnetic structure does not align with any of the Q-SDW ordering candidates as reported in \cite{wang2024electronic}.
Since we focus on the large $J$ limit that corresponds to the high-pressure phase, these differences in magnetic structures potentially underscore the significance of the change of point groups symmetry and enhanced inter-layer exchange $J$ in the magnetic correlation during structural transition, though $y$-direction ferromagnetic solution might results from the artifact of a too small width dimension. 

The relative phase of the magnetization for the $3d_{x^2-y^2}$ and $3d_{z^2}$ orbitals on each atomic site is indeed influenced by the sign of the next-nearest-neighboring inter-orbital hopping integral. Specifically, when orbitals hybridize through a positive $t_{x^{2}-y^{2},z^{2}}$, they share the same sign of spin, while a negative $t_{x^{2}-y^{2},z^{2}}$ results in orbitals sharing opposite signs of spin. 

Since our analyses essentially are based on the plaqutte ladder geometry, whether this spin density structure preserves in the two-dimensional limit requires further studies in the future. It is natural to find that the spin structure is different from previous study for the pure one-dimensional setup \cite{shen2023effective} due to the change of system dimensions.

\begin{figure}[t]
\includegraphics[width=0.5\textwidth]{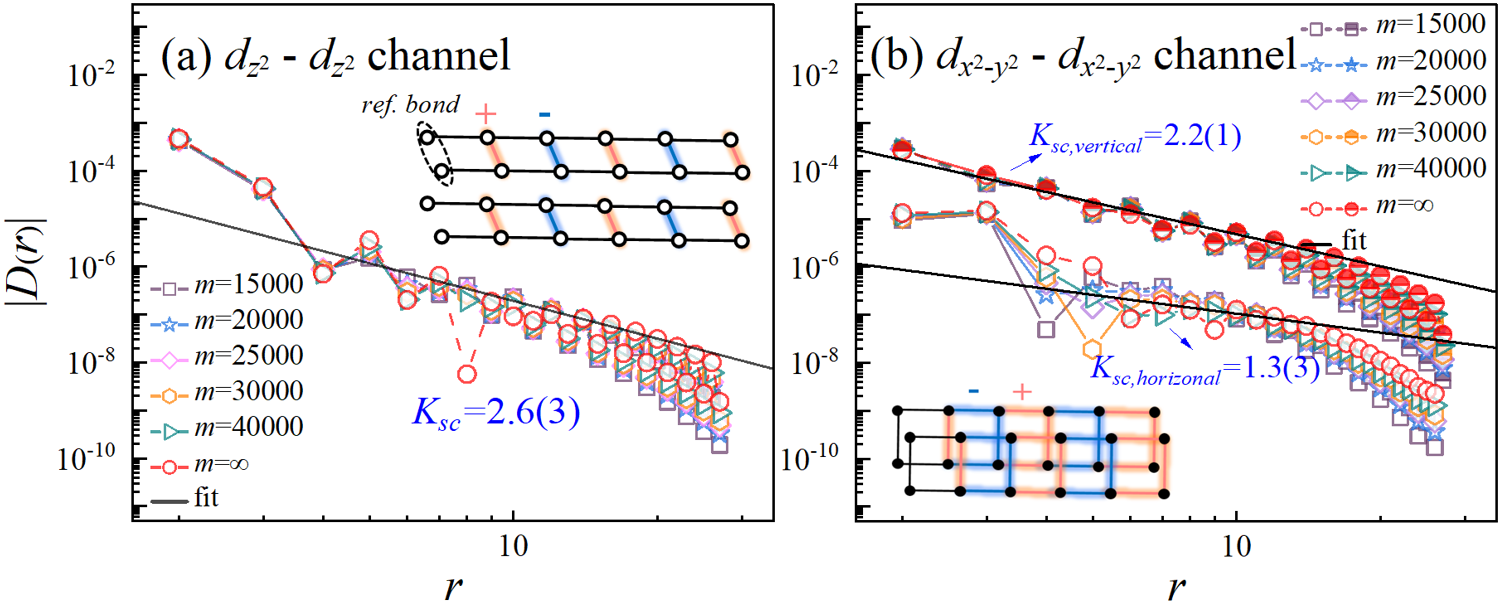}
\caption{The spin-singlet pair-pair correlation functions for (a) $3d_{z^{2}}-3d_{z^{2}}$ and (b) $3d_{x^{2}-y^{2}}-3d_{x^{2}-y^{2}}$ channels. The inter-layer $3d_{z^{2}}$ bonds at $x = 4$ are set as the reference bonds (see insets). The absolute values of the pair-pair correlation $|D(r)|$ are plotted on a double logarithmic scale. Both results of finite kept states and extrapolation are shown. The envelope fitting gives an algebraically decay with power-law exponents $K_{sc}$.  The long-distance pair-pair correlation decay for all these channels has a periodic oscillations as spin density, and the corresponding sign structures are plotted in the insets.}
\label{pair}
\end{figure}

\textbf{Superconducting correlations}. - In order to study the superconducting instability and the pairing symmetry, we calculate the
equal-time spin singlet superconducting pair-pair correlation function $D(i,j)=\langle \hat{\Delta%
}_{i}^{\dagger }\hat{\Delta}_{j}\rangle $, where
\begin{equation}
\hat{\Delta}_{i}^{\dagger }= (\hat{c}_{(i,1),\uparrow }^{\dagger }\hat{c}%
_{(i,2),\downarrow }^{\dagger }-\hat{c}_{(i,1),\downarrow }^{\dagger }\hat{c}%
_{(i,2),\uparrow }^{\dagger })/\sqrt{2}.
\end{equation}
Different pairing channels are calculate by setting bond $i$ (formed by site $(i,1)$ and $(i,2)$) and bond $j$ (formed
by site $(j,1)$ and $(j,2)$) to different orbitals.
We set the 3$d_{z^{2}}$ bond at $x=4$ (the dashed oval in Fig.~\ref{pair} (a)) as the reference bond in the
calculation of the pair-pair correlations to make sure the pairing correlation results are least influenced by the boundary \cite{PhysRevB.108.165113}. In Fig.~\ref{pair} (a), we show the results for the $3d_{z^2}$ channels. A polynomial fit of the decaying of absolute value with power-law function $|D(r)|\sim r^{-K_{sc}}$ gives exponent $K_{sc} = 2.6(3)$. Interestingly, the sign of the pairing correlation oscillates with period $2$ in the $x$ direction. The similar fit gives $K_{sc} = 1.3(3)$ and $2.2(1)$ for horizontal and vertical $3d_{x^2-y^2}$ bonds, as shown in Fig.~\ref{pair} (b)). 
Same as the $d_{z^{2}}$-$d_{z^{2}}$ channel, the sign of the pairing correlation also oscillates with period $2$ in the $x$ direction for the $d_{x^{2}-y^2}$-$d_{x^{2}-y^2}$ channel. Considering the sign-changing oscillations, pairing correlation can be described by $D(r)\sim r^{-K_{sc}}cos(Q \cdot r+\theta)$, where ordering vector $Q=\pi$ and $\theta$ is the phase shift. Given $\chi _{sc}(T)\sim T^{-(2-K_{sc})},\text{ when }T\rightarrow 0$, the pairing for both channels are close to or within the region with divergent pairing susceptibility. 

In the bi-layer two-orbital model in Eq.~(\ref{ham}), the pairing driving force results from the inter-layer AFM super-exchange of the $d_{z^{2}}$ orbitals. But without the hybridization with the $d_{x^{2}-y^2}$, the system could not display superconductivity due to the absence of coherence of the paired electrons \cite{shen2023effective,PhysRevB.108.L201108}. Similar as the one dimensional case \cite{shen2023effective}, pairing correlation is stronger in the $d_{x^{2}-y^2}$ channel, even though  $d_{x^{2}-y^2}$ orbital is not the source of pairing.  We notice that the pairing correlation are weaker than the one-dimensional system \cite{shen2023effective}. The oscillation of the sign for both channels indicate the possible pair density wave \cite{agterberg2020physics} in the system, different from the either the $s_{\pm}$ \cite{PhysRevB.108.L140505,gu2023effective,luo2024high,PhysRevLett.131.236002} and $d$-wave \cite{luo2024high,liu2023role,Jiang_2024,PhysRevB.110.024514} pairing symmetry determined in previous works. 
But whether the sign structure of pairing persists in the two dimensional requires further study of larger systems. 

{\textbf{Discussion and Conclusion}.} -
{In the atmoic limit when inter-orbital hybridization $t_{x^{2}-y^{2},z^{2}}$ is absent, two orbitals are decoupled. The ground state of
the two-site half-filled $3d_{z^{2}}$ orbitals forms an isolated singlet in the ground state due to the antiferromagnitc exchange $J$, while the ground state of the interacting $3d_{x^{2}-y^{2}}$ falls into a uniform charge state with gapped excitations \cite{PhysRevB.107.125114} on two-leg Hubbard ladders. With finite $t_{x^{2}-y^{2},z^{2}}$, 
when the filling of $3d_{x^{2}-y^{2}}$ orbitals is slightly away from $1/2$, and the $3d_{z^{2}}$ bound pairs can effectively hop on the $x-y$ plane through the itinerant $3d_{x^{2}-y^{2}}$ channels. The emergence of superconductivity necessitates the phase coherence of these pre-formed bound pairs.
} 

In this work, we focus on the large $J$ region and tune the difference of chemical potential to target $1/16$ doping in the $d_{z^{2}}$ orbital. However, if the chemical difference is fixed, the increase of $J$ (corresponding to the increase of pressure experimentally) could decrease the doping in the $d_{z^{2}}$ orbital and shrink the size of $\gamma$ pocket, which is harmful for superconductivity. This can explain the experiment results \cite{li2024pressure} that when the pressure is increased to $18$ Gpa, $T_c$ starts to decrease and superconductivity terminates at $90$ Gpa. 

In conclusion, we have conducted a study on the bi-layer two-orbital model for pressured La${_3}$Ni${_2}$O${_7}$ on a plaquette ladder, employing large-scale DMRG calculations with meticulous extrapolation of truncation errors.  Our findings reveal that the properties of this model on a plaquette ladder differ from those previously studied in minimal one-dimensional configurations \cite{shen2023effective}. We observe Neel-type quasi-long-range antiferromagnetic correlations for both the $3d_{x^{2}-y^{2}}$ and $3d_{z^{2}}$ orbitals. Additionally, the $3d_{x^{2}-y^{2}}$ orbitals exhibit quasi-long-range CDW order with a wavelength of 2, highlighting the orbital-selective nature of charge ordering. Notably, sign oscillations in pairing correlations are detected for both orbitals, which contrasts with previous studies \cite{PhysRevB.108.L140505,gu2023effective,luo2024high,PhysRevLett.131.236002,luo2024high,liu2023role,Jiang_2024,PhysRevB.110.024514}. {The presence of orbital-selective CDW, quasi-long-range Neel type antiferromagnetic correlations and possible pair density wave instability underscores the complex interplay of multi-orbital nature and various electronic correlations in this material.}  Although our investigation is limited to a plaquette ladder rather than a true two-dimensional lattice due to computational constraints, the insights gained into spin, charge, and pairing correlations are invaluable for elucidating the properties of La${_3}$Ni${_2}$O$_{7}$.    

\textbf{Acknowledgments}. - Y. Shen and M. P. Qin thank Weidong Luo
for his generosity to provide computational resources for this work. M. P. Qin
acknowledges the support from the National Key Research and Development
Program of MOST of China (2022YFA1405400), the National Natural Science
Foundation of China (Grant No. 12274290), the Innovation Program for Quantum Science and Technology (Grant No. 2021ZD0301900), and the sponsorship from Yangyang
Development Fund. G. M. 
Zhang acknowledges the support from the National Key Research and
Development Program of MOST of China (Grant No. 2023YFA1406400). The calculation in this work is carried out with TensorKit \cite{foot}. 

\bibliography{main}
\end{document}